\documentclass[11pt]{article}
\input epsf.sty
\usepackage{epsfig} 
\newcommand{\beq}{\begin{eqnarray}}
\newcommand{\eeq}{\end{eqnarray}}

\newcommand{\M}{M}
\newcommand{\Mt}{{\cal M}}

\newcommand{\be}{\begin{equation}}
\newcommand{\ee}{\end{equation}}
\newcommand{\ben}{\begin{eqnarray}\displaystyle}
\newcommand{\een}{\end{eqnarray}}
\newcommand{\refb}[1]{(\ref{#1})}

\newcommand{\sectiono}[1]{\section{#1}\setcounter{equation}{0}}

\newcommand{\Tr}{{\rm{Tr}}}

\newcommand{\Z} {{\bf Z}_2}

\def\sqr#1#2{{\vcenter{\vbox{\hrule height.#2pt
         \hbox{\vrule width.#2pt height#1pt \kern#1pt
            \vrule width.#2pt}
         \hrule height.#2pt}}}}

\begin{document}

{}~ \hfill\vbox{\hbox{hep-th/0410121} \hbox{PUPT-2132}

\hbox{HUPT-04/A036}

 }\break

\vskip 1cm

\begin{center}
\Large{\bf Minimal Superstrings and Loop Gas Models}

\vspace{0.6cm}


\vspace{20mm}

\large{Davide Gaiotto$^{a}$ , Leonardo Rastelli$^{b}$
and Tadashi Takayanagi$^{a}$}

\vspace{10mm}

\large{\em $^{a}$ Jefferson Physical Laboratory,
Harvard University,}

\vspace{0.2cm}
\normalsize{\em Cambridge, MA 01238, USA}

\vspace{0.4cm}

\large{\em $^{b}$ Joseph Henry Laboratories, Princeton University,}

\vspace{0.2cm}

\large{\em Princeton, New Jersey 08544, USA}

\vspace{0.5cm}

\end{center}

\vspace{10mm}

\begin{abstract}

\bigskip

We reformulate the matrix models of
minimal superstrings  as loop gas models on random surfaces.
In the continuum limit, this leads to the
identification of minimal superstrings with certain bosonic string theories,
to all orders in the genus expansion.  RR vertex
operators arise as operators in a $\Z$ twisted sector
of the matter CFT. We show how the loop
gas model implements the sum over spin structures
expected from the continuum RNS formulation.
Open string boundary conditions are also more transparent in this language.

\medskip

\end{abstract}

\newpage

\tableofcontents

\section{Introduction }

String theories in low dimensions have recently come back into focus as
simplified but still very instructive laboratories for string physics.
The bosonic theories were originally solved  through the double-scaling
(=continuum) limit of hermitian matrix models, which  implement a discretization
of the bosonic worldsheet \cite{DS}-\cite{GZ}
The modern understanding of D-branes, open/closed duality, and tachyon
condensation, together with crucial technical advances in Liouville CFT 
\cite{DornOtto}-\cite{FH}, 
have inspired a reinterpretation of the double-scaled matrix
model as the  ``open string field theory on an infinite number of decayed localized
D-branes''
 \cite{MV,KMS}. In turn,
this insight has led to the identification of the matrix models corresponding to
non-critical  {\it super}string theories ($\hat c = 1$ Type  0B  in \cite{chat1, chat2}, 
$\hat c = 1$  Type 0A in \cite{chat2} and $\hat c <1$ 0A and 0B  in \cite{chat3}), 
and to a renewed interest in the subject ({\it e.g.} \cite{renewedfirst}-\cite{renewedlast}).

\smallskip

The Type 0B cases are solved by unitary matrix models \cite{GW, PS, CDM0, CDM, BDJT, HMPN}
(equivalently hermitian models with a double-cut ansatz),
while the Type 0A cases are solved by complex matrix models \cite{DJM,  LM}. 
This is surprising, 
as these matrix models are close cousins of the hermitian models that yield the bosonic 
string, and
it  is not at all clear in which sense (if any) unitary and complex models are providing
a discretization of {\it super}Riemann surfaces. 
This is not just a conceptual puzzle. These matrix models provide exact answers even 
in the presence of RR fluxes,  while worldsheet CFT techniques would seem as problematic as in the usual
 critical string case.
Connecting the matrix models with the continuum worldsheet formulation
we can hope to learn something about the worldsheet realization of RR backgrounds.

\smallskip

In this paper we make some progress in this direction.  We focus on the $(2, 4k)$ sequence of minimal
superstrings,  which are obtained by coupling $(2, 4k)$
superminimal models to superLiouville theory and are related to one-matrix models \cite{chat3} --
but we expect our approach to generalize to all minimal superstrings. 
The main technical novelty is to reformulate these matrix models in terms of 
a gas of self-avoiding fermionic loops 
on a random surface \cite{KoL}. 
The random surface is the discretization of a bosonic worldsheet,
with the loop gas as additional matter degrees of freedom. 
This implies that minimal superstrings 
are perturbatively  equivalent to certain {\it bosonic} string theories,
to each order in the genus expansion. 
The loop gas formulation allows to read off  quite directly the matter CFT of these bosonic string theories.
For example, in the simplest case of the $(2,4)$ model (pure supergravity, or $\hat c = 0$)
the matter theory is the free $c_M = -2$ CFT of two Grassmann odd scalars $\Theta^+(z, \bar z)$ and
$\Theta^- (z, \bar z)$. 

\smallskip

A general feature is that NSNS vertex operators map to vertex operators 
in the bosonic string with ``integer'' Liouville dressing;
 while RR vertex operators   have ``half-integer'' Liouville dressings, their matter part 
belongs to a $\Z$ twisted sector of the matter CFT and they introduce cuts in the bosonic worldsheet,
in analogy with the continuum RNS formulation. On  higher genus surfaces, the loop gas model correctly 
implements the sum over spin structures, again as expected from the RNS formulation.  
Thus, while we are formulating minimal superstrings (in the genus expansion) as certain bosonic string models, 
we seem to have got one step closer to the fermionic RNS formulation.
It is an intriguing open question whether one can complete
this program and recover  our loop gas models from the point of view
of  discretizing surfaces with a metric {\it and} a gravitino.

\smallskip

Open strings boundary conditions seem also more transparent in this framework. 
We find indications that the loop gas 
language accommodates naturally both ``electric'' and ``magnetic''  branes ($ \eta = \pm 1$ in the notations of \cite{SS}), while
 only the $\eta = -1$ branes are  transparent in the one-matrix model \cite{chat3, SS}. We find that the $\eta = -1$ branes are related
to Dirichlet boundary conditions for the fermionic matter degrees
of freedom $\Theta^\alpha$ arising from the loop gas; 
they correspond to the standard resolvant of the matrix model.  
We propose that the $\eta = +1 $ branes are related instead to  Neumann boundary conditions for the $\Theta^\alpha$
fields, and  correspond to a novel type of resolvant in the matrix model. 

\smallskip

For the closed string sector of the 0A models, our claims
reduce to well-established results \cite{chat3, cliffold, cliff1, cliff2}. 
0A  minimal superstrings with positive cosmological constant 
have no local RR vertex operators in the closed string sector \cite{chat3}.
The $(2, 4k)$ 0A theory is perturbatively the same as the minimal $(2, 2k-1)$ bosonic string 
\cite{chat3, cliff1, cliff2}. The  0B models are obtained by taking a $\Z$ orbifold of the 0A models 
(and vice-versa), where the $\Z$ acts as $-1$  $(+1)$ on all RR (NSNS) vertex operators. 
Thus one may {\it a priori} expect the $(2, 4k)$ 0B model to be described perturbatively by a $\Z$ orbifold of the
minimal $(2, 2k-1)$ bosonic string. The loop gas formulation
makes this relation transparent, clarifying how exactly the orbifold works both in the
matrix model and in the continuum worldsheet formulations.

\smallskip

The detailed organization of the paper can be gleaned from the table of contents.
Sections 2, 3 and 4 deal with the closed
string sector.  Section 2 introduces
the basic setup. Loop gas models
for 0A (0B) theories are deduced by
gauge-fixing complex (two-cut hermitian) 
 matrix models; the 0B version is related to 0A version by a $\Z$ orbifold
that introduces a bicoloring of the random surface.
Section 3 deals in detail with the $\hat c= 0$ case,
where an interesting matrix model with Parisi-Sourlas
supersymmetry \cite{KMK, Dav, KM,  KW1, KW2, EdwardsKlebanov, plefka}
plays an important role. Section 4
treats the continuum worldsheet formulation
of minimal superstrings as bosonic strings.
Section 5 focusses on D-branes,
both in the discretized and in the continuum formalisms.
Section \ref{conclusions} contains some concluding remarks.
An appendix reviews properties
of the $c_M = -2$ ``symplectic fermions'' CFT \cite{Kausch1,  Kausch2,  Flohr, GK1, GK2, KoganWheater, 
KawaiWheater, BredthauerFlohr}, and  details some of its uses in connection with minimal superstrings.

\sectiono{The superstring from the  loop gas}
\label{superstring}

We begin in this section  by considering the Type 0B $(2, 4k)$ theories.
We take the cosmological constant $\mu$ to be positive
since our goal is to make contact with the 
continuum worldsheet formulation, and large positive $\mu$
 is the usual perturbative regime.\footnote{The matrix model results
 suggest that large negative $\mu$ corresponds to {\it another} perturbative regime,
but it is not obvious how to reconstruct the corresponding continuum worldsheet theory.}
These theories are solved by hermitian matrix models with a double-cut ansatz \cite{chat3}.

\smallskip

Recall that hermitian one-matrix model is defined by computing correlators
with the measure 
\be 
\label{onematrix}
\int d\M \,  e^{-\frac{1}{g  }\,{\rm Tr}
V(\M)} \, , \ee 
where $\M$ is an $N \times N$ hermitian matrix. The
model is  solvable by reducing the matrix integral to an
$N$-dimensional integral over the eigenvalues, which can then be
evaluated using orthogonal polynomials theory (for a review see
{\it e.g}. \cite{DGZ}). All correlators of basic $U(N)$ invariant
observables, {\it e.g.} $Tr \M^k$ and ${\rm det}(x-\M)$, can be {\it exactly}
evaluated this way. Alternatively, one can set up a perturbative
expansion around a semiclassical vacuum. To expand around a
one-cut vacuum (all eigenvalues in one potential well) it is
convenient to shift $\M$ so that the bottom of the well is  at $\M =
0$, $V'(\M=0)= 0$. Treating cubic and higher terms of the potential
as perturbations gives the usual expansion in ribbon Feynman
diagrams, which can be organized according to their genus. Viewing
these diagrams as discretized Riemann surfaces with a metric leads
to the connection with minimal bosonic strings. 

\smallskip

Recently, the two-cut vacua of the hermitian one-matrix model for
the left-right symmetric potential $V(\M)=V(-\M)$
have been connected  with Type 0B minimal superstrings 
with  $\mu > 0$ \cite{chat3}.  The ${ \Z}$ 
symmetry of the potential can be explained from the viewpoint
of the open string field theory in the superstring.
At present there is no connection between this matrix model and the viewpoint
of discretizing superRiemann surfaces with a metric {\it and} a gravitino.

\subsection{Gauge fixing and the two-color loop gas}

\label{gaugefixing}

To understand what kind of worldsheets this matrix model
is discretizing, we now set-up a perturbative Feynman diagram expansion around the two-cut vacuum. 
It is natural to partially gauge-fix the
action making $\M$ block-diagonal,
$\M = {\rm diag} (\M_{L}, \M_{R})$,  and then expand independently
in terms of $\M_L$ and $\M_R$ around the two different critical points.

\smallskip

The measure factor for the change of variables from $\M$ to $\M_L$
and $\M_R$ is standard (see {\it e.g.} 
\cite{BDE, DGKV}.)  The Jacobian  determinant is the determinant
of the linear action
\be \label{linear}
\sigma   \to [\sigma,{\rm diag}(\M_L,\M_R)] \, ,
\ee
 where $\sigma$ is a  block off-diagonal matrix. Hence
\be \label{dM2}
d\M = {\rm det}( \M_L \otimes I_{N_R} - I_{N_L} \otimes \M_R)
{\rm det}( \M_{R} \otimes I_{N_L} - I_{N_R} \otimes \M_{L})
\, d\M_L \, d\M_R \, d \Sigma \, ,
\ee
where $\Sigma$ parametrizes $\frac{U(N)}{U(N_L) \otimes U(N_R)}$ and
$I_K$ is the $K \times K$ identity matrix. It is useful to rewrite this determinant as an integral over ghost variables,
two Grassmann-odd  $N_L \times N_R$  matrices $C_1$, $C_2$ and their
$N_R \times N_L$ conjugate matrices $\bar C_1$, $\bar C_2$. The model \refb{onematrix}
becomes 
\be
\int d \M_L\, d \M_R \, d C_1 \, d \bar C_1 \, d 
C_2   \, d \bar C_2 \; e^{- S_{2}} 
\ee
where
\be
\label{ActionTwo}
S_{2} \equiv\frac{1}{g} {\rm Tr}[V(\M_L)  +  V(\M_R)  + \bar C_1 \M_L C_1  - \bar C_2 \M_L C_2  - \M_R
\bar C_1 C_1  +  \M_R \bar C_2 C_2  ] \,.
\ee
Assume now that $V(\M)$ is symmetric under $\M \to -\M$ and
has a minimum at $\M =a$ and $\M=-a$. It is convenient
to define $\M_L \equiv - a + \Phi_L $ and $\M_R \equiv  a - \Phi_R$,
 \be \label{signs}
 \M  = \left( \begin{array}{cc}  -a + \Phi_L   &  0  \\  0  &    a - \Phi_R  \end{array}\right)\, ,
\ee
so that the two critical points correspond to $\Phi_L = 0 $
and $\Phi_R = 0$. With this shift the ghosts acquire a kinetic term,
\ben 
\label{Two}
S_{2} & =&  \frac{1}{g} \, {\rm Tr} [ V(\Phi_L - a) + V(\Phi_R - a) 
 - 2 a (\bar C_1 C_1 -  \bar C_2 C_2)  \\
&& \quad  \quad+ \bar C_1 \Phi_L C_1   -\bar C_2 \Phi_L C_2 
 + \Phi_R \bar C_1 C_1 -  \Phi_R  \bar C_2 C_2  ] \,.\nonumber \een
 In analyzing this matrix model, we should remember
 that we are interested only in observables
 which are gauge-invariant with respect to the full $U(N)$ symmetry
 of the original one-matrix model \refb{onematrix}, for example 
 \be \label{rro}
  {\rm Tr}\, \M^n = \Tr(a-\Phi_R)^{n} + (-1)^n \Tr(a-\Phi_L)^{n}\,. 
 \ee
 If we add these operators to the action, we change the $L$ and $R$ potentials in a correlated way, corresponding to an analytic deformation of the 
 original one-matrix model potential $V(\M)$. Keep also in mind that because
 of our choice of signs in \refb{signs}, operators which are even (odd) under $ \Phi_L \leftrightarrow \Phi_R$
 correspond to even (odd) powers of the original field $\M$.

 \begin{figure}
\centering \epsfig{file=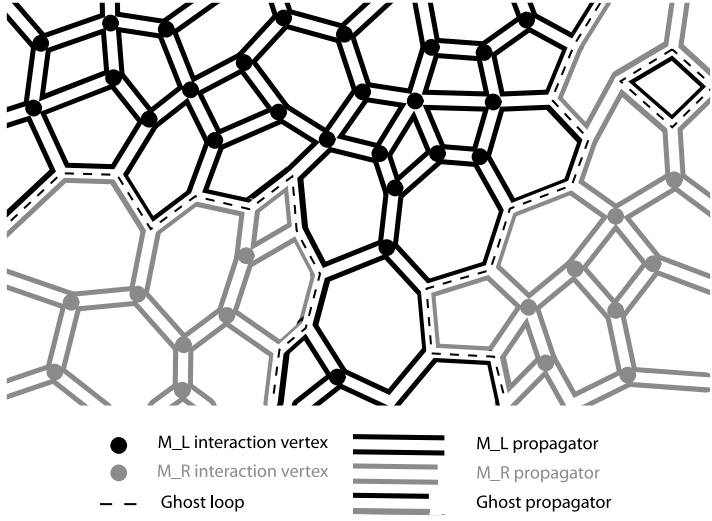, width=4in}
\caption{A Feymann diagram for the two-color $O(-2)$ matrix model
\refb{Two}.
\label{bicolor}}
\end{figure}
\smallskip
 \smallskip
 
The action \refb{Two} admits a good perturbative expansion.
Feymann diagrams are easy to picture. Since the ghosts appear only quadratically, 
their propagators form self-avoiding closed loops in the Feymann diagram.
There are no vertices mixing directly $\Phi_L$ and $\Phi_R$,
which interact only through the coupling with the ghosts.
The ghosts are $N_L\times N_R $ matrices so in the usual
double-line notation each ghost loop has two
sides, on one side only $\Phi_L$ legs are inserted, on the other only $\Phi_R$.  Hence any Feymann graph will consist of patches of two different ``colors",
$L$ or $R$, separated by the ghost loops (see Figure \ref{bicolor}).
The dual diagrams are random surfaces with the extra degrees of freedom of
a gas of self-avoiding loops.  There is a constraint on
configurations, as there has to be an assignation of $L$ or $R$
color to each region bounded by a loop, with {\it different} colors on the two sides of each loop.
In the evaluation of the diagram,  each self-avoiding loop carries
a weight of $-2$ (the minus sign because of odd-Grassmanality,
and the factor of two because there are two copies $i=1,2$ of the
$C_i$, $\bar C_i$)\footnote{It is clear that
despite the sign differences in the action, the two flavors
of ghosts play a symmetric role.  A $C_2$  ghost loop
has a relative factor of $(-1)^{V_L + V_R +P}$ with respect
to a $C_1$ ghost loop, where $V_L$ and $V_R$ are
the numbers of $L$ and $R$ vertices in the ghost loop
and $P$ the number of ghost propagators. Since $P = V_R + V_L$,
this factor is one. One could of course redefine variables to have $C_1$ and $C_2$
appear symmetrically in the action, but our choice is more natural in 
relation to the one-color model.}.

\smallskip

Loop gas models on random surfaces are well studied  ({\it e.g.}  \cite{KoL, DK, K2, KS, EZ, EK, KazKostov, KostovBoundary, Knew}). 
 In the $O(n)$ model one assigns a weight of $n$ to each self-avoiding loop.
Our case is then related to the $O(-2)$ model.  However the usual version of the 
$O(-2)$ model \cite{KoL} is formulated on a random surface with only ``one-color''. 
It is immediate to write the  matrix model whose perturbative expansion 
generates the one-color  $O(-2)$ loop gas \cite{KoL},
\be
\int d \Phi \, d C \, d \bar C \; e^{-S_1}\, , 
\ee
where
 \ben
 \label{One}  S_1 & \equiv & \frac{1}{g} \, {\rm Tr} [ V(\Phi - a)  - 2 a \bar C C  +  \bar C  \Phi C  + \Phi \bar C C]
 \\
& = & \frac{1}{g} \, {\rm Tr} [ V(\Mt)  +  \bar C  \Mt C  + \Mt \bar C C]
\, .\nonumber
\een
(Here we have defined $ \Mt \equiv \Phi - a$.) We obtain the two-color model \refb{Two} 
from \refb{One} by setting
\be \label{projection}
\Phi = \left( \begin{array}{cc}   \Phi_L  & 0 \\   0  & \Phi_R \end{array} \right)\,, 
\quad C = \left(\begin{array}{cc}   0 & C_1 \\ \bar C_2  & 0 \end{array} \right) \,.
\ee 
The two-color model can be thought of as a $\Z$ orbifold
of the one-color  model.   The orbifold  operation introduces a twisted sector of
operators {\it odd} under $\Phi_L \leftrightarrow \Phi_R$.  At genus zero 
the untwisted sector (operators {\it even} under $\Phi_L \leftrightarrow \Phi_R$)
is isomorphic to the one-color model, since ``coloring'' is trivial on the sphere.  At higher genus
the two theories  differ also in the untwisted sector, since  the two-color model has an additional constraint on 
configurations:  the number of ghost loops linked to each non-trivial homology cycle of
the surface must be even.

\subsection{Twist lines and RNS}

\label{RNS}

A familiar way to implement  an orbifold is to relax the constraint
and introduce ``twist lines": On a surface of genus $g$, we pick $2g$
independent  homology cycles ${\cal C}_j$, $j=1, \dots, 2g$,
and write the partition function as an average
over the $2^{2g}$ sectors labeled by the choices
of signs  $\epsilon_j = \pm 1$  for each homology cycle.
Introducing into the partition function the factor of $ \prod_j \epsilon_j^{\# {\rm loops \; linking \;}{\cal C}_ j}$
implements the ${\bf Z}_2$ orbifold.\footnote{One says that the sector
with $\epsilon_j = -1$ {\it has }a twist line along ${\cal C}_j$,
while the sector with $\epsilon_j = +1$ does {\it not} have a twist line.}
 For example, for the genus one vacuum amplitude, we can write 
\be \label{torusa}
Z_{g=1}=\sum_{k,l\in {\bf Z}} \frac{(1+(-1)^k)(1+(-1)^l)}{4}
Z^{(0)}_{g=1}(k,l)= \sum_{k,l\in 2{\bf Z}} 
Z^{(0)}_{g=1}(k,l)
\, , \ee where the integers $k$ and $l$ denotes the
number of ghost loops homotopic to the $a$- and $b$-cycle of the torus,
respectively; we also defined the amplitude $Z^{(0)}_{g=1}(k,l)$
of the matrix model without any color constraint. The sector that
includes the factor $(-1)^k$, for example, corresponds to the
torus with the twist-line along  the $b$-cycle.

\smallskip

\begin{figure}
\centering \epsfig{file=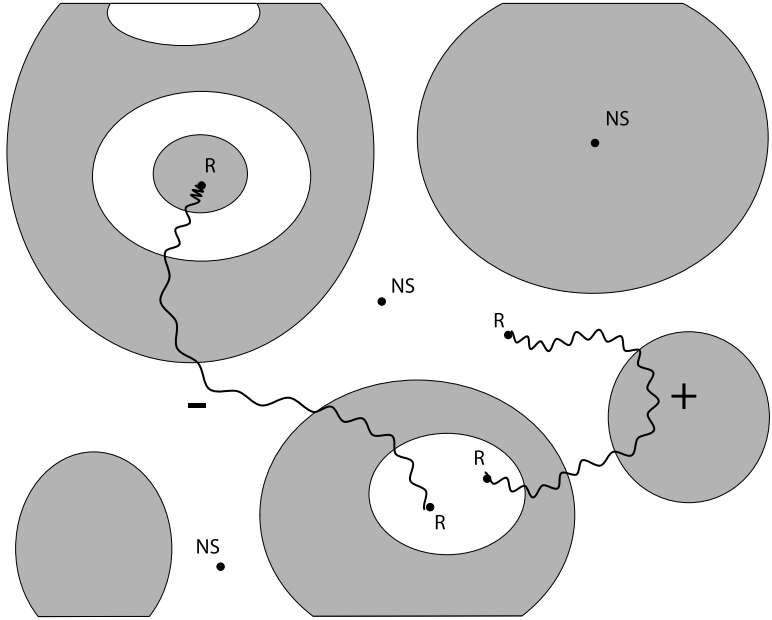, width=4in} \caption{Twist
lines and Ramond operators \label{twistline}}
\end{figure}

Matrix model operators consisting of even (odd) power of 
$\M$ are even (odd) with respect to the symmetry $\Phi_L
\leftrightarrow \Phi_R$ (see \refb{rro}).  In the formulation where we drop the color constraint and add twist lines,
the appropriate signs for odd operators 
are accounted for by making them end-points for twist lines. Indeed a twist line
between two insertion points will give a minus sign if they lie in patches of different color, a plus if they lie in 
patches of the same color (see Figure \ref{twistline}).
In the double-scaling limit of the two-cut matrix model, the even
and odd operators correspond respectively to NSNS and RR vertex
operators \cite{chat3}.  It is then compelling to identify the $2^{2g}$ sectors described
above with the $2^{2g}$ choices of spin structures in the
continuum RNS  formulation.  For example, the torus amplitude
(\ref{torusa}) can be regarded as that projected by the  Type 0 GSO
projection \be
Z_{g=1}=\Tr_{NSNS}\left[\frac{1+(-1)^{f_L+f_R}}{2}\right]
+\Tr_{RR}\left[\frac{1+(-1)^{f_L+f_R}}{2}\right], \ee where
$f_{L,R}$ is the left or right-moving world-sheet fermion number.
Twist lines joining odd operators in the loop gas model
clearly correspond to  cuts joining spin fields on the RNS worldsheet.

\smallskip

This seems a step forward in understanding how the matrix model is providing
a discretization of a superRiemann surface. Clearly this program is not complete, since the two-color matrix
model \refb{Two} is building {\it bosonic} random surfaces, with some extra matter degrees
of freedom in the form of ghost loops. It would be very interesting to derive \refb{Two} from a super-geometric formulation
that includes explicitly the path integral over the gravitinos.
What we have learnt so far is that the minimal Type 0B strings must have (at least
perturbatively) a  formulation as bosonic strings.

\subsection{Relation with 0A}
\label{0A}

With some hindsight, the perturbative equivalence of  
the 0B theory to a bosonic string is  not too surprising.  The $\Z$ orbifold relating
the two-color to the one-color model projects out all
odd operators. In the continuum limit, 
this is  the orbifold relating the 0B to the 0A minimal superstrings, 
which projects out all RR closed string  vertex operators.  It is known that the string equations for 0A $(2, 4k)$ model\footnote{
For $q=0$, of course. More on the meaning of $q$ momentarily.} 
are perturbatively equivalent to  the string equations for the bosonic $(2, 2k-1)$ models \cite{cliffold, chat3, cliff1, cliff2}. 
Hence we should expect that the $(2, 4k)$ 0B theory is perturbatively equivalent
to a $\Z$ orbifold of the bosonic $(2, 2k-1)$ model.

\smallskip

This circle of ideas can be completed by showing 
directly that the complex matrix model for the 0A theory 
is equivalent to the one-color loop gas.\footnote{We thank
N. Seiberg for  explaining this to us.} The complex matrix model reads
\be
\int d \Mt d \Mt^\dagger     e^{-\frac{1}{g}  {\cal V} (\Mt^\dagger  \Mt)   }  \, ,
\ee
where $\Mt$ is a complex $N \times (N+q)$ matrix. Taking $q=0$,
we can gauge-fix to $\Mt = \Mt^\dagger$ hermitian and positive-definite.
By a reasoning analogous to (\ref{linear},\ref{dM2}), 
we are led to introduce the ghosts  $C$, $\bar C$ and 
to reproduce precisely the action \refb{One}.   In deriving the one-color model by this route, the 
integration in \refb{One} is restricted to  positive-definite matrices $\Mt$,
\be \label{One>}
\int_{\Mt \geq 0} \,
 d {\Mt}
\, d C \, d \bar C \; \exp \left( -\frac{1}{g} \, {\rm Tr} [ V(\Mt)  +  \bar C  \Mt C  + \Mt \bar C C] \right)
\,  ,
\ee 
where $V(\Mt) \equiv {\cal V}(\Mt^2)$. This restriction makes no 
difference to each order in the perturbative expansion, and corresponds to 
the natural non-perturbative definition of the  0A model.  Let us
mention for completeness the generalization 
to  $q \neq 0$.  The gauge-fixing requires the introduction of some extra fermionic variables, 
the $q \times N$ matrix $\bar b $  and the $N \times q$ matrix $b$,
leading to
\be \label{One>q}
\int_{\Mt \geq 0} \,
 d {\Mt}
\, d C \, d \bar C \; \, d b \, d \bar b \;\exp \left( -\frac{1}{g} \,\left( {\rm Tr} [ V(\Mt)  +  \bar C  \Mt C  + \Mt \bar C C]
+ \bar b  \Mt^2  b  \right)  \right)
\,  .
\ee

\smallskip

The perturbative equivalence between the $(2, 4k)$ 0A with $q = 0$ and the
$(2, 2k-1)$ bosonic models \cite{cliffold, chat3, cliff1, cliff2} can be readily
understood by establishing a map between the respective matrix models, as
we now review.
Diagonalizing $\Mt = {\rm diag}(\mu_i)$ and integrating out the ghosts in \refb{One>}, one has 
\be \label{map}
\prod_{k=1}^N 
\int_{0}^\infty d \mu_k  \,    \prod_{i < j}   (\mu_i - \mu_j)^2  \, \prod_{i, j}  (\mu_i + \mu_j)\, e^{-\frac{1}{g}  V(\mu_k)  }\,.
\ee
The change of variables $h_i = \mu_i^2$ gives
\be \label{H}
\prod_{k=1}^N \int_0^\infty d h_k  \, \prod_{i < j}   (h_i - h_j)^2 \, e^{-\frac{1}{g}  V( \sqrt{h_k} )} = \int_{H \geq 0}  dH   
\, e^{ - \frac{1}{g} V(\sqrt{H})}\,.
\ee
This is almost the same as the hermitian one-matrix model for the matrix $H \equiv \Mt^2$,
with the differences that 1)For generic potentials $V(\Mt)$, the action
contains square roots of $H$; 2)The integration runs only over 
{\it positive} eigenvalues of $H$.  
However, for our purposes, we are only interested in {\it even} polynomial potentials $V(\Mt)$,
 which are  polynomials in $H$. This is obvious from the point
 of view of the 0A model, since $V( \sqrt{H}) = {\cal V} (H)$.  Following the route from the 0B theory,
 in going from the two-color to the one-color
model  we are restricting by definition to the ``untwisted'' sector of operators
even under $\Phi_L \leftrightarrow \Phi_R$, which map to even powers of $\Mt$, ${\rm Tr} \Mt^{2n} = {\rm Tr} H^n$.
  Moreover,  the difference in the integration region has only effects which are non-perturbative in $1/N$,
and is thus immaterial to all orders in the genus expansion.  Thus the one-color model  is (perturbatively) equivalent to
 the standard  one-matrix model. Its multicritical points\footnote{multicritical polynomial potentials
in $H$ of degree $k+1$, descending from multicritical even potentials $V(\M) $ of degree  $2k+2$ in
the original two-cut model \refb{onematrix}}  
correspond in the continuum limit to  the  $(2, 2k-1)$ Virasoro minimal models coupled to Liouville,
as anticipated.

\sectiono{ Loop gas models for pure supergravity}

\label{24}

We now elaborate on the simplest example. The $(2,4)$  0B theory ($\hat c = 0$)  is obtained
from the two-cut matrix model with a quartic potential $V(\M) = \frac{1}{4}(\M^2 - a^2)^2$ \cite{chat3}.
In the large $N$ limit, for $\mu > 0$ one considers a two-cut ansatz for the eigenvalues.
The critical point  is reached when the two cuts meet at the origin with a zero of the 
resolvant $R(z)$. 
In our normalizations, this happens for $g  N  \to  \frac{a^4}{4}$ from below.  The double-scaling limit consists in 
setting $g N =\frac{ a^4}{4} - \epsilon \, \mu $,  $\epsilon  \to 0$, where $\mu$ is the bulk cosmological constant, and
zooming in around the origin of the $z$ plane, $z   = \epsilon \,  x$.

\smallskip

In the loop gas model (either in the two-color or one-color version), this critical
point corresponds to the so-called ``dilute phase''. The loop degrees of freedom are becoming massless
as we approach the critical point  (the effective mass for the ghosts is the distance of the
eigenvalue cut from the origin), so that the
number of vertices occupied by the ghosts is diverging.
Simultaneously the potential $V$ is also becoming critical, 
which causes the number of unoccupied vertices to also diverge. (By contrast, 
the dense phase would correspond to the ghost
degrees of freedom filling the random surface in the continuum
limit.) In fact, this particular choice of quartic potential corresponds
to a higher multicritical point, where both cubic {\it and} quartic
vertices are diverging \cite{KS}.
Standard formulas for the $O(n)$ model (see {\it e.g.} \cite{Knew}) 
give the the central charge of the matter and Liouville continuum CFT,
\be \label{nb}
n = - 2 \cos \left( \frac{\pi}{  b^2}  \right) \,, \quad  Q = b + \frac{1}{b}  \, ,\quad 
 c_L = 1 + 6 Q^2 =  26 - c_M  \, ,   
\ee
where  $b <1$   ($b >1 $) in the dilute (dense) phase.\footnote{We are using conventions
where the {\it boundary} cosmological constant operator is $e^{b \phi}$. 
The {\it bulk } cosmological constant operator is always $e^ { 2 b_{-} \phi} $ where $b_{-}$ is the smaller
of $b$ and $1/b$. Bulk operators must always obey the Seiberg bound,
boundary operators need not.} In our case we have $n=-2$ in the dilute phase,
and the multicritical quartic potential corresponds to $b = 1/\sqrt{2}$,
 $c_M = -2$, $c_L = 28$, which are indeed the values expected for the minimal $(1, 2)$ bosonic string
theory.\footnote{In principle $n=-2$ can be obtained by $b^2= 1/ (2m)$ for any integer $m$,
but as shown in \cite{KS} the case of multicritical quartic potential corresponds to $b= 1/\sqrt{2}$.}
Notice that we are landing on theory with  $b = 1/\sqrt{2}$ rather than $b = \sqrt{2}$,
so we are accurately referring to this theory as the $(p, q ) = (1,2)$ model (as opposed to $(2,1)$),
in conventions where $b^2 = p/q$. This subtlety is  immaterial
for the closed string sector, which is our focus in this section, but plays
a role for D-branes, as we shall see in section \ref{Dbranes}.

\smallskip

Following (\ref{map},\ref{H}),
the one-color loop gas model with quartic potential $V(\Mt)$ can be mapped to a one-matrix integral
with a gaussian potential, 
\be \label{Hgauss}
\int_{H \geq 0}  dH \, e^{ - \frac{1}{4 g}  (H-a^2)^2} \,.
\ee
In this model, the large $N$ eigenvalue density is the usual Wigner distribution, 
supported in the interval $[ \frac{a^2}{2} - (g N)^\frac{1}{2} ,  \frac{a^2}{2} + (g N)^\frac{1}{2}]$. 
As  $g N \to \frac{ a^4}{4}$, the left boundary of the cut reaches
the origin (which is also the boundary of the integration region
in \refb{Hgauss}). The double-scaling limit is again
obtained by zooming in around the origin. The double-scaling limit
of the gaussian model  is well-known to yield the $(2,1)$ minimal bosonic string \cite{GM, KW2}, 
which is equivalent to topological gravity \cite{LPW, Wit, distler, DVVL},
see \cite{MMSS} for a recent simple derivation.\footnote{The gaussian model is accurately labeled as $(2, 1)$ (as opposed to $(1, 2)$). More in section
\refb{etam} on the distinction between $(1, 2)$ and $(2, 1)$.} Our model is not quite the usual gaussian model in that we have a restriction of the integration region to positive $H$
 eigenvalues. We can think of having an infinite potential wall at $ H=  0$ that prevents
 eigenvalues to explore the region with $H < 0$.  In the doubled-scaled
variables, the wall is a finite distance away in units of $\mu$  and its effects are non-perturbative in  $g_s$.

\subsection{The supersymmetric matrix model }
\label{susy}

The case of quartic potential $V$  is especially interesting since we can view
the one-color model \refb{One} as a  matrix model with Parisi-Sourlas
supersymmetry.  Introducing the matrix superfield
\be
{\bf \Phi}(\theta, \bar \theta) = \Phi + \bar \theta C + \theta \bar C
+ \theta \bar \theta F \, ,
\ee
we can rewrite action $S_1$ in superspace \cite{Dav, KM},
\be \label{Ssuper}
S_1  =\frac{1}{g} \, \int d \theta d \bar \theta \; {\rm Tr}
( \partial_\theta {\bf \Phi} \partial_{\bar \theta} {\bf \Phi}
+ {\cal W} ( {\bf \Phi}) )\, , \quad
{\cal W}({\bf \Phi}) = - a {\bf \Phi}^2 + \frac{1}{3} {\bf \Phi}^3 
\,.
\ee
In fact, this supersymmetric model is equivalent to the gaussian model 
\refb{Hgauss} for {\it any} choice of superpotential ${\cal W} (\bf \Phi)$, by the
``Nicolai map''  $H = {\cal W}' (\Phi)+a^2$ \cite{KM}.  This formulation
has  also an immediate interpretation as a theory of matrix fields living in minus two
dimensions \cite{KMK,Dav,KM}.  The perturbative expansion in superspace realizes the discretization of a Riemann 
surface embedded in minus two dimensions, which of course agrees with the value $c_M = -2$ of the continuum string theory.
 The Feynman diagrams in superspace have
 a $\int d  \theta_j d \bar \theta_j $ integral at  vertex $j$
and a propagator $1/a^2 -  (\theta_j -\theta_k)(\bar \theta_j -
\bar \theta_k)/a$. The propagators can be collected in an exponential like
\be e^{-a \sum_{(jk)} (\theta_j -\theta_k)(\bar
\theta_j - \bar \theta_k)} \ee
This is the discretization of the continuum CFT of two free Grassmann odd scalars, with action 
\be
\label{Stheta} \frac{1}{4 \pi}  \int d^2 z  \, \epsilon_{\alpha \beta}
\partial \Theta^\alpha
\bar \partial \Theta^\beta \, ,
\ee
where $\alpha = \pm$. The superspace coordinates correspond to the zero modes of the $\Theta^\alpha$
fields, $\theta = \theta_0^+$, $\bar \theta = \theta_0^-$. 
We refer to the appendix for more details on this CFT, which
plays an important role in the following.

\smallskip

We can now  describe the continuum
version  of the ${\bf Z_2}$ orbifold that leads
from the  one-color to the two-color loop gas model.
The effect of a twist line in the discretized theory is to
flip the sign of the ghost propagators that cross it.
This corresponds in the continuum to flipping the sign
of the $\Theta^\alpha$ across the twist line, that
is, to performing the orbifold $\Theta^\alpha \to - \Theta^\alpha$.
 We conclude that $(2,4)$ minimal fermionic string is equivalent to
a bosonic non-critical string defined by coupling to Liouville
the matter CFT with $c_M = -2$ obtained by
the $\Z$ orbifold  of the  $\Theta^\alpha$ system.

\subsection{Open string field theory}
\label{OSFT}

Finally we are in the position to point out a transparent open string field theory interpretation 
of the one- and two-color matrix models,  in the same  spirit as  the proposal of \cite{MV,KMS,MTV}
for the $c=1$ matrix model. Starting with the one-color model, we can regard \refb{Ssuper} as the effective open string
field theory on infinitely many localized D-branes of the $(1, 2)$ minimal bosonic
string. We consider D-branes which have ZZ boundary conditions in the Liouville direction and
Neumann boundary conditions for the $\Theta^\alpha$ CFT. Since the coordinates
$\theta, \bar \theta$ that represents the positions of the D-brane are fermionic,  
we get the supermatrix model (\ref{Ssuper}). Incidentally, it is amusing to recover
the matrix action as the action of Witten's cubic OSFT \cite{OSFT}
(on  $N$ ZZ branes with Neumann
boundary conditions for the $\Theta^\alpha$), gauged-fixed
to Siegel gauge and  truncated to the lowest modes. Expanding the open string field as
\be
|\Psi\rangle = \sum_{i, j = 1}^N \left(
 M_{ij} \,  c_1 | 0 \rangle_{ij} +   C_{ij} \,  \theta^-_0 \,  c_1 | 0 \rangle_{ij} +  \bar  C_{ij} \, \theta_0^+ c_1 | 0 \rangle_{ij}
 + F_{ij} \theta^+_0   \theta^-_0 \, c_1 | 0 \rangle_{ij} + \dots \right) \, ,
 \ee
 one can evaluate the Witten action with the usual CFT rules \cite{LPP}. The
 kinetic term $\langle \Psi, c_0  L_0 \Psi  \rangle$
 gives precisely the quadratic terms in \refb{Ssuper}, with
 the $\partial_\theta \partial_{\bar \theta}$ derivatives
 arising from the zero mode terms of $L_0$; while the
 interaction term $\langle \Psi, \Psi, \Psi \rangle$ gives the
 cubic term of the superpotential.

 \smallskip

The $\Z$ projection  \refb{projection} leading to the two-color model \refb{Two} can
now be recognized as the usual rule \cite{DM} to perform   a ${\Z}$ orbifold 
on the D-brane worldvolume,
\be
{\bf \Phi}(\theta, \bar \theta)=\sigma_3\cdot {\bf \Phi}
(-\theta, -\bar \theta)\cdot \sigma_3 \, ,
\ee
where $\sigma_3 = {\rm diag}  (I_{N_L}  , - I_{N_R} ) $.

\sectiono{The continuum bosonic formulation }

\label{continuum}

The previous arguments show that we can view the
$(2, 4k)$ minimal superstrings as
bosonic strings, to each order in perturbation theory. The 0A theories map to the bosonic $(2, 2k-1)$ strings \cite{chat3, cliffold, cliff1, cliff2},  
while the  0B theories to a $\Z$ orbifold of these models. 
We now analyze this correspondence in the continuum worldsheet formulation.

\subsection{The operator spectrum}
\label{operatorspectrum}

The first task is to establish a map between the physical vertex operators 
of the superstring and the physical vertex operators of the bosonic
formulation. We shall see that the bosonic CFTs 
 are not the usual  $(2, 2k-1)$ {\it minimal} models (at least not for the 0B cases),
 but extensions with an infinite number of Virasoro primaries.
A complete understanding will require the specification of the precise
operator content and the construction of modular invariant
partition functions. Here we consider the simpler problem of listing the 
Virasoro representations that appear in these CFTs. Our discussion partly overlaps
with \cite{chat3, cliff1}. In this subsection we assume $k > 1$ and come back to the special case $k=1$ below.

\smallskip

The primary operators  of the   $(p, q) = (2, 4k)$ superminimal models 
are labeled in the usual Kac table notation as $(r, s) = (1, s)$, with  $1 \leq s \leq 2 k$.
Operators with odd $s$ are in NSNS sector, while operators
with even $s$ are in the RR sector. The operator $(1, 2 k)$ is the RR ground
state, while $(1,1)$ is of course the identity.  In the 0A theory with $\mu > 0$,
only NSNS operators survive the GSO projection, while in the 0B theory
with $\mu > 0$ all operators are kept.  To construct the physical
tachyon vertex operators  ${\cal T}_{rs}$ one needs to gravitationally dress the superconformal
primaries with a Liouville factor\footnote{Here we are using the same conventions as \cite{SS}. The background charge is given by
$Q=b + 1/b$, $b = \sqrt{p/q}$ and the Seiberg bound is at $Q/2$,
for both the fermionic and models. The central charge of the Liouville sector is
${\hat c}_L=1+2Q^2$ for superLiouville and $c_L=1+6Q^2$ for bosonic Liouville.}
 $e^{\beta_{rs} \phi}$,
with the exponent obeying the Seiberg bound $\beta_{rs}  \leq Q/2$. It 
is convenient to quote the rescaled dressings $\gamma_{rs}=   \beta_{rs}/Q $
(the rescaling is such that the Seiberg bound is at 1/2). For the
$(2, 4k)$ models,
\be \label{gammaf}
\gamma_{1\,s} = \frac{s + 1}{2(2k+1)} \,. 
\ee

\smallskip
In the bosonic $(2, 2k-1)$ minimal models, operators are labeled
as $(\tilde r, \tilde s ) = (1, \tilde s)$ with $1 \leq \tilde s \leq k-1$.
Tachyon vertex operators $\tilde {\cal T}_{1\, \tilde s}$
have rescaled Liouville dressings (again in units where the Seiberg bound is at 1/2)
\be \label{gammab}
\tilde \gamma_{1 \, \tilde s} = \frac{  2\tilde s + 2 }{2(2k + 1)}\,.
\ee
We shall shortly need the more general formula
\be \label{gammag}
\tilde \gamma_{\tilde r \, \tilde s} = \frac{2 k+1 - \tilde r (2k-1)  +  2\tilde s  }{2(2k + 1)} \, .
\ee
The  matching between the bosonic and fermionic
theories is now straightforward using KPZ scaling.
KPZ scaling \cite{KPZ,DDK} tells us that the partition function $F_g(\{t_n \})$ at genus $g$, as a 
function of the sources $\{ t_n \}$ for the local operators of the theory, is
a homogeneous function of degree $Q(1-g)$, where each $t_n$ is
assigned the weight  $\gamma_n$ equal to its rescaled Liouville (or superLiouville)
dressing. Comparison of \refb{gammaf} and \refb{gammab} leads to $s =2 \tilde s + 1$,
{\it i.e.} to the identification 
\be
 {\cal T}_{1 \,  2 \tilde s +1}  \cong \tilde {\cal T }_{1 \, \tilde s } \,.
\ee
As expected, the bosonic vertex operators map to NSNS vertex operators in the superstring.
Notice that the operator ${\cal T}_{11}$, which corresponds to  
the cosmological constant deformation of superLiouville theory, is absent in the minimal
bosonic theory. It maps in the bosonic theory to $(1, \tilde s = 0)$,  
which is just outside the minimal Kac table.  This is not surprising, since
this operator  is associated with a redundant deformation in the KdV hierarchy --
which to each order in perturbation theory governs the 0A theory as well.
There are $(k-1)$ independent deformations in the integrable hierarchy, and
$k$ NSNS operators in the superstring;  the $k$-th operator (the one with lowest superLiouville dressing) must be redundant. 
In the bosonic theory this operator has a Liouville dressing equal to one half the Liouville dressing of the bulk
cosmological constant (which has  $\tilde s = 1$, of course) 
and it can be interpreted as a ``boundary operator'' \cite{boundary}. In modern language, it corresponds
to the boundary cosmological constant operator, which can be inserted on boundaries with FZZT boundary
conditions.

\smallskip

For the 0A theory this is the end of the story. For the 0B theory, we
need to find the counterpart of the RR operators on the bosonic side. Inspection
of \refb{gammaf}  and \refb{gammag} leads to match the superconformal
labels $(1, s)$ for $s$ even with the bosonic labels $(\tilde r = 2, \tilde s)$
with $s = 2 ( \tilde s - k+1)$,
{\it i.e.} to the identification
\be \label{2s}
 {\cal T}_{1 \,  2 (\tilde s+ 1 -k)}  \cong \tilde {\cal T }_{2 \, \tilde s } \, ,
\ee
with $  k  \leq  \tilde s \leq 2k-1 $.
Clearly all of these operators are outside the minimal bosonic
Kac table. Closure of the fusion rules will
generically require to include {\it all}
degenerate Virasoro representations $(2, \tilde s) $ and
$(1, \tilde s)$ with no restrictions on $\tilde s$.
In the 0B case the enlargement of the Kac table is inevitable, 
and this suggests it may be natural to treat the 0A case  using a non-minimal table as well
(see \cite{nonminimal} for some relevant discussions).
One can check that the bosonic fusion rules  are compatible with the expected $\Z$ symmetry, 
with the operators $(\tilde r, \tilde s)$  being even for $\tilde r = 1$ and odd for $\tilde r = 2$. 
It would be interesting to construct modular invariant partition functions
for these non-minimal conformal field theories.\footnote{There is some
work on non-minimal $c = 0$ theories, which are related
to percolation \cite{percolation} and the $SU(2)$ WZW model at level zero \cite{su20}.}

\subsection{${\bf \hat c =0}$ from ${\bf c = -2}$ }
\label{continuum24}

The case of the bosonic $(2, 1)$ model (more accurately $(1, 2)$), 
for which $c_M = -2$,   requires a separate treatment.  This theory has been extensively studied  \cite{Kausch1,  Kausch2,  Flohr, GK1, GK2, KoganWheater, 
KawaiWheater, BredthauerFlohr}
as the simplest example
of a logarithmic CFT.  There is a free field representation
of the model in terms of a pair of Grassmann odd scalars $\Theta^+ (z, \bar z)$ and  $\Theta^- (z, \bar z)$. 
Modular invariant partition functions have been constructed,
for various choices of operator content corresponding to various orbifolds of the  $\Theta^\alpha$ fields. 
Although these partition functions contain an infinite number of Virasoro primaries, they are rational
with respect to a $W$-algebra.  It is conceivable that a similar story could be worked out for the other 
$(2, 2k-1)$ models.
In the appendix we collect some useful formulas pertaining to the $\Theta^\alpha$ system.  

\smallskip

The bosonic theory corresponding perturbatively to the $(2, 4)$ 0A model
can be  defined taking as matter CFT simply the  free $\Theta^\alpha$ system with periodic
boundary conditions.  The  natural tachyon vertex operators are \cite{GR}
\be \label{Vn}
{\cal O}_{n}=e^{\frac{3-n}{\sqrt{2}}\phi}{\cal P}_n(\partial \Theta^\alpha, \bar \partial \Theta^\beta)\, c \bar c\, , \quad
n = 1, 3, 5, \dots
\ee
Here $\phi$ is the Liouville field and $c$, $\bar c$ the 
usual reparametrization ghosts of the bosonic string, and
${\cal P}_n$ are matter primaries of dimension $(n^2 -1)/8$
 built acting on the vacuum with $\Theta^\alpha$ oscillator.  In the notation
of the previous section, these primaries correspond to the $(1, \tilde s)$ Virasoro representations, with $\tilde s = (1-n)/2$.
 The primary ${\cal P}_{n = 1}$ is just the identity, its Kac label being $(1, 0) \equiv (1,1)$
 by the usual reflection property $(\tilde r, \tilde s ) \equiv (p-\tilde r, q - \tilde s)$.
 ${\cal O}_1$ is thus the bulk cosmological constant
 operator, and it maps to the only tachyon operator of the 0A $(2,4)$
 model, namely the cosmological constant operator of the superLiouville
 theory ($s = 1$ in the notations of the previous section). In this case the cosmological
 constant deformations of the fermionic and bosonic  string theories are in correspondence with each other:
 the ``conformal'' background of the bosonic $(1, 2)$  
model maps to ``superconformal'' background  
of the $(2, 4)$ model. This is not the case for the models with $k >1$.

\smallskip

For $n >1$, the operators \refb{Vn} are interpreted as ``gravitational descendants''
in the topological gravity language. From the viewpoint of the bosonic string theory, 
they are standard elements of the cohomology at ghost number $(1,1)$. Gravitational
descendants are often represented as elements of the cohomology at non-standard ghost numbers \cite{LZ},
in fact this is necessarily the case if the matter CFT is truly minimal.
In our case, the matter CFT has infinitely many primaries and we gain
the ability to represent gravitational descendants as standard vertex operators
of ghost number $(1,1)$, which is perhaps a more convenient picture for actual
calculations.  This phenomenon is not new and has been discussed in the topological string theory
literature in related contexts, see for example \cite{BLNW, losev, lerche}.

\smallskip

The theory corresponding to the $(2, 4)$ 0B model is obtained  by the $\Z$ orbifold
$\Theta_\alpha \to - \Theta_\alpha$.  The orbifold introduces a twisted sector built
with $\Theta^\alpha$ oscillators acting on a vacuum of dimension $-1/8$.
We refer to the appendix for the expression of the torus partition
function. The twisted sector can be decomposed in irreducible Virasoro representations
of dimensions $(n^2 -1)/8$, where now $n$ is even. All these
Virasoro representations are also degenerate and correspond
to the $(0, \tilde s)$ elements of the Kac table, with $\tilde s = -n/2$. 
 Tachyon vertex operators are obtained by dressing with the
same Liouville exponential as in \refb{Vn}, with $n$ even.
The tachyon with $n = 0$ maps to the RR ground state
of the $(2,4)$ model.  Tachyons with  $n>0$ are again
interpreted as gravitational descendants.

\subsection{The torus path-integral}

As a check of our proposal, we can evaluate the torus partition function in the continuum bosonic
formulation of the $(2, 4)$ model
and compare it with the matrix model
result. The partition function in the
continuum {\it fermionic} formulation of the $(2, 4k)$ models
 has not yet been fully computed, because  the odd spin structures are somewhat subtle.
 The results for the even spin structures are available \cite{BK}, 
and they match with the matrix model results \cite{chat3}.
For the 0A theory with $\mu > 0$, the spectrum has only
NSNS operators, the torus partition function coincides with that of the bosonic $(2, 2k-1)$
theory,  and we do not have anything new to add to the discussion in \cite{chat3}. 
For the $(2, 4)$ 0B theory with $\mu > 0$  we should
be able to compute the full torus partition function using our identification with the
$\Z$ orbifold of the $(1, 2)$ model.
The expected matrix model result is (for general $k$ and $\mu > 0$)
\be \label{Z0B}
Z_{2 ,4k}^{0B} = - \frac{2k+1}{24 k} \, \log  t  \, ,
\ee
where $t$ is the coupling of the operator
of lowest dimension  in the theory.

\smallskip
This calculation is straightforward using the
results of \cite{BK}. These authors computed
the string theory torus path integral in the continuum formulation; 
in the case that the matter CFT  is a boson at radius $R$  they found
\be \label{ZR}
Z ^{c=1}(R/\sqrt{\alpha'})   = -\frac{1}{24}  \left( \frac{R}{\sqrt{\alpha'}}  + \frac{\sqrt{\alpha'}}{R} 
\right) \log t \,.
\ee
As reviewed in the appendix,
the $\Z$ orbifold of the $\Theta^\alpha$ system
has a partition function identical to a compact boson of radius $R = \sqrt{2\alpha'}$.
Thus the calculation of the torus path integral is  almost identical to the one that leads
to \refb{ZR} with $R = \sqrt{2 \alpha'}$. The only
difference is in the integral over the Liouville zero mode,
which contributes the factor of  $- \log {t}/{\beta}$,
where $\beta$ is the Liouville dressing of the most relevant matter operator.
The torus path integral $Z_{\Z}$ of the $\Z$ orbifold of the $(1, 2)$ model
can then be deduced from \refb{ZR} if we account
for the different values of $\beta$. The most
relevant operator of the $(1, 2)$ theory is the
cosmological constant, which in our normalizations
has $\beta_{c = -2} = \sqrt{2}$; the most
relevant operator for $c=1$ is again the cosmological
constant, which has $\beta_{c=1} = 2$. Hence 
\be
Z_{\Z} = 
 \frac{\beta_{c=1}}{\beta_{c=-2}} \,   Z^{c=1}(\sqrt{2 } ) 
 =   -\frac{1}{8} \,\log t \,  .
\ee
This agrees with the matrix model result \refb{Z0B}\footnote{
Curiously, the matrix model result \refb{Z0B} equals for all $k$ 
(up to an overall factor of 1/4) the continuum calculation
of \cite{BK} for the minimal bosonic strings with modular
invariant partition functions of type $A_{p-1   } D_{1+ q/2}$, with the
formal substitution $p =2 $, $q =2k-1 $. 
 This suggests a connection between the $\Z$ orbifold relating 0A and 0B and
the $\Z$ orbifold relating the $AA$ and $AD$ bosonic
theories, although the connection is not straightforward
because here we are dealing with non-minimal CFTs.}
 for $Z^{0B}_{2 , 4}$.

\sectiono{D-branes}

\label{Dbranes}

A very interesting comparison involves the realization of open string
boundary conditions in the loop gas model (and its continuum limit)
versus the continuum fermionic formalism.  
The spectrum of FZZT branes in the continuum fermionic
formulation 0A and 0B theories has been determined in \cite{SS}. Let
us recall their results.
 In the  $(2,4)$  0B theory with $\mu > 0$,
there are three relevant FZZT  boundary states
\cite{FZZT,TE, FH}, 
 labeled  as \cite{SS}\footnote{
We  are omitting the dependence on the label $\zeta\equiv {\rm sign} (\mu) $ of \cite{SS}
since we  restrict our analysis to $\mu > 0$,
so that $\zeta \equiv 1$,  $\hat \eta  \equiv \zeta \eta = \eta$.}
\be  \label{Bbranes}
| x;  \eta= + 1  \rangle_B  \,,    \qquad | x;   \eta= -1 \, , \xi = \pm 1 \rangle_B \, .
\ee
Here $x$ is the boundary cosmological constant,
which takes values in the auxiliary Riemann surfaces ${\cal M}_{2, 4}^{ \eta = \pm }$ introduced in 
\cite{SS}.
The label $\eta = \pm 1$ distinguishes the linear
combination of left and right moving supercharges
annihilating the boundary state
(``electric'' versus ``magnetic'' branes)
while $\xi = \pm 1$ distinguishes
a brane from its anti-brane.  
The $| x;  \eta= + 1  \rangle$ brane does not couple to the RR ground state
and it is thus identical to its own anti-brane.\footnote{
However for the $(2, 4k)$ 0B models with $k >1$, 
the $\eta = +1$ brane does couple to RR fields and thus
the distinction $\xi = \pm 1$ is relevant.}  Moreover,
computations of one-point functions indicate
that in BRST cohomology one can identify $| x; \eta= - 1; \xi  \rangle = 
|- x ;  \eta= - 1; =\xi  \rangle$,
and parametrize the independent $\eta = -1$ branes just by their value of $x$, 
forgetting the label $\xi$.
Finally from the annulus amplitude, one finds that the open strings
between branes with the same $\eta$ are in the NS sector,
while open strings between branes with opposite $\eta$
are in the Ramond sector. Results for the higher  $(2, 4k)$ 0B models
with $\mu > 0$ are similar. A priori the
boundary states would acquire labels corresponding to the matter primaries, however it is believed \cite{SS} (in analogy
with the bosonic string) that in the BRST cohomology such labels are redundant
and one can still parametrize the inequivalent boundary states
by  $\eta = \pm 1$, $\xi= \pm 1$ and $x$ taking
value in the appropriate Riemann surface ${\cal M}^{\eta = \pm }_{(2, 4k)}$.

\smallskip

 On the other hand,  in the $(2,4)$ 0A theory with $\mu > 0$, the relevant boundary states are
\be \label{Abranes}
| x;  \eta= + 1 \, , \xi = \pm 1 \rangle_A  \,,    \qquad | x;   \eta= -1 \,  \rangle_A \, .
\ee
In the 0A theory with $\mu > 0$ the RR ground state
does not give rise to a local vertex operator \cite{chat3}, but it
corresponds to the zero mode of the RR field, under which the $\eta = + 1$ brane is charged;
changing the sign of $\xi$ reverses the charge of the brane.

\smallskip
\begin{figure}
\centering \epsfig{file=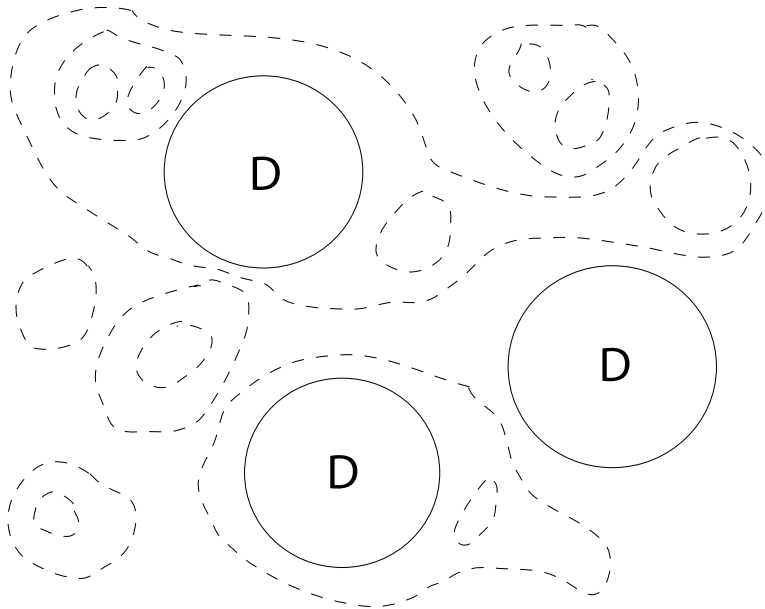, width=4in}
\caption{Dirichlet boundary conditions for $\Theta^\alpha$
 in the loop gas model.  The dashed lines represent the
 self-avoiding ghost loops.  
 \label{DirBrane}}
\end{figure}

\subsection{The ${\bf \eta = -1}$ brane}
\label{etam}

For both 0A and 0B, the FZZT branes with $\eta = -1$ have been identified \cite{chat3, SS}
with the  standard macroscopic loop operators of the corresponding matrix model.

\subsubsection{0A}

Let us consider the 0A case first. The loop operator is\footnote{As usual, the operator creating a
brane is actually the exponential of the macroscopic loop operator $W(x)$.
 Also, by a slight abuse of notation we are using the same letter $x$ in the matrix model operator
$W(x)$ and  in the continuum  boundary state $|  x;  \eta = -1 \rangle$,
while the two are of course related by the standard renormalization.} 
\be \label{WA}
W^{\eta = -1}_A (x_A) = - \frac{1}{N} {\rm Tr \; log} ( \Mt^\dagger \Mt- x_A)  
\to -\frac{1}{N}  {\rm Tr \; log} ( \Mt^2 -x_A) = 
 -\frac{1}{N} {\rm Tr \; log} ( H- x_A)     \, ,
\ee
where the arrow indicates the gauge-fixing procedure of section \ref{0A}.
In terms of the matrix $H$, this is the standard resolvant
of the one-matrix model; perturbatively, results
for the resolvant expectation values in $(2, 4k)$ 0A coincide with those in  $(2, 2k-1)$ bosonic.
However it would be wrong to attempt to match the auxiliary
Riemann surfaces computed in \cite{SS} between bosonic
and fermionic models. This is because the results of
\cite{SS} are valid for conformal or superconformal
backgrounds, but for general $k$ the cosmological
constant deformation of the bosonic theory  does not correspond to the
cosmological constant deformation of  the 0A theory.
They do match for $k=1$; indeed in this case
the $\eta  = -1$, $\mu > 0$ curve for 0A coincides with the
curve $ 2y^2 -1 = x $ \cite{SS} of the $(2, 1)$ bosonic theory 
(up to rescalings of $x$ and $y$).

\smallskip

It is clear that in our formulation in terms of the one-color loop gas,
 $W^{\eta = -1} _A(x)$ cuts a macroscopic hole which {\it cannot} 
 intersect any of the self-avoiding ghost loops, since the matrices $C_i$ do not
appear in $W^{\eta=-1}_A$ (see Figure \ref{DirBrane}).   Let us now
focus on the $k=1$ case.  In the supersymmetric formalism of section \ref{susy},  
we can write
\be
W^{\eta = -1}_A (x_A) =  - \frac{1}{N} {\rm Tr \; log} ( ({\bf  \Phi  } -a)^2  -x_A) |_{\theta = \bar \theta = 0}\,.
\ee
In the continuum limit, this operator cuts a hole in the worldsheet
with {\it Dirichlet} boundary conditions $\Theta^\alpha = 0$ for the $c_M = -2$ CFT. Notice however that
in the supersymmetric model, the more natural expression for the resolvant is
\be
W_{(1, 2)} (x ) = - \frac{1}{N} {\rm Tr \; log} ( ({\bf  \Phi  } -a  -x) |_{\theta = \bar \theta = 0}\,.
\ee
In the double-scaling limit, $W_{(1,2)}$ and $W^{\eta = -1}_A (x_A)$ 
are equivalent with  $x_A $ scaling as $ x^2$. This relation between $x$ and $x_A$ reflects
the fact that the continuum limit of the one-color gas model gives what should
be more accurately referred to as the $(p, q ) = (1, 2)$ minimal
bosonic string, the one with  $b \equiv \sqrt{p/q} = 1/\sqrt{2}$;
while the continuum limit of the gaussian model gives
the $(2, 1)$ theory with $b = \sqrt{2}$. There
is no distinction between $(1, 2)$ and $(2, 1)$ in the closed string sector
but open string boundary conditions are labeled differently. 
The boundary cosmological $\mu_B$ constant  of the $(1, 2)$ model is
equal  to the {\it dual} boundary cosmological $\tilde \mu_B$
of the $(2, 1)$ model,   and vice-versa.
Now $\mu^{(2,1)}_B \equiv \tilde \mu^{(12)}_B = 2 (\mu^{(1,2)}_B)^2 -\mu$,
which is compatible with the identifications $x_A \sim \mu_B^{(2,1)}$ and
$x \sim \mu_B^{(1,2)}$.  FZZT branes with Dirichlet
boundary conditions for $\Theta^\alpha$ in the $(1,2)$ have been considered
in \cite{GR}, where is is shown that the full cubic OSFT
on them localizes to the Kontsevich model. The authors of
 \cite{MMSS} reproduced this result starting with the resolvant \refb{WA} 
for the gaussian matrix model in terms of the matrix $H$, and performing the double-scaling limit;
they observed that in their procedure one lands to the $(2, 1)$ model
with $b = \sqrt{2}$, a fact that we have just explained from
 the relation between the gaussian and loop-gas matrix models.\footnote{
The correspondence between the gaussian matrix model
and the $(2, 1)$ string theory can also be deduced using
the general loop gas formula \refb{nb}. Besides the dilute
phase realization with $n = -2$ and $b = 1/\sqrt{2}$, there
is a dense phase realization of $c_M = -2$, with $n = 0$ and $b = \sqrt{2}$.
Since $n = 0$, we have just the one-matrix model, and since we are in
the dense phase, the potential is {\it not} at a critical point -- and
we may as well take it to be gaussian without changing
universality class.}

\smallskip
More generally the full superfield loop operator, 
\be
W_{(1, 2)} (x,  \rho^\alpha )  =
 - \frac{1}{N} {\rm Tr \; log} ( ({\bf  \Phi  } -a  -x)|_{\theta = \rho,  \bar \theta = 
 \bar \rho}\, ,
\ee
gives in the continuum limit the FZZT brane tensored
with the Dirichlet boundary state 
$| {\rm  Dir}(\rho^\alpha) \rangle$
 for the $\Theta^\alpha $ system (see \refb{Dgen} in the appendix).

\subsubsection{0B}

Let us now consider the $\eta = -1$ brane for the 0B model.  This is naturally
identified with the macroscopic loop operator
of the two-cut matrix model,
\ben
\label{Wx}
W^{\eta = -1}_B(x_B)&  =&- \frac{1}{N} {\rm Tr \; log} ( M- x_B)  = -\frac{1}{N}
{\rm Tr \; log} ( -x_B -a + \Phi_L)
-\frac{1}{N}  {\rm Tr \; log} (- x_B + a - \Phi_R)
\nonumber \\
&= & \frac{1}{N} \, \sum_l  \frac{ {\rm Tr} \,  \Phi_L^l}{l \, (a +x_B)^l}  +
\frac{1}{N} \, \sum_l  \frac{ {\rm Tr} \,  \Phi_R^l}{l \, (a-x_B)^l}  -\frac{1}{N} 
\log(x_B^2 - a^2) \,.
\een
Again in the two-color loop gas model the ghost loops
do {\it not} intersect the boundary (Figure \ref{DirBrane}).
 It is also clear from \refb{Wx} that the expansion
of $W^{\eta = -1}_B(x_B)$ contains both even and odd operators under $\Phi_L \leftrightarrow \Phi_R$, 
which in the continuum limit are identified as NSNS and RR operators.
This is just as expected for the $\eta = -1$ branes, which
couple both to NSNS and RR closed operators (see \refb{Bbranes}).
We can be more precise.
 In the expansion of $W^{\eta = -1}_B(x_B)$, we  weight a boundary of length $l$
differently according to its color, either with a factor proportional
to $1/(a+x)^l$ for the color $L$ or with a factor proportional
to $1/(a-x)^l$ for the color $R$.  To express this fact in the
formalism of the unconstrained loop gas with twist lines,
we must extend to sum to new sectors, allowing for twist
lines that end on a boundary. The weight for a boundary of length
$l$ must be taken
\be
\frac{1}{l \, (x_B+a)^l} + \frac{\epsilon}{l \, (x_B-a)^l}
\ee
where $\epsilon = +1$ in the sector where no twist
line ends on the boundary and $\epsilon = -1$ if it does.
This fits perfectly with the continuum RNS formulation, where the $\eta = -1$ branes have both an RR and an NSNS
term in their boundary state and thus one must sum
over the different sectors in the closed channel.
If we remove the distinction between $L$ and $R$ (that is, project
out the odd operators), 
\be
W^{\eta = -1}_B (x_B) \longrightarrow
-\frac{1}{N}  {\rm Tr \; log} ((\Phi-a)^2- x_B^2)=
W^{\eta = -1}_A (x_A = x_B^2) \, .
\ee
The relation $x_A = x_B^2$ is as expected \cite{chat3, SS}.

\smallskip

We can now analyze the 0B $\eta = -1$ brane in the continuum
bosonic formulation. The boundary state is obtained by
taking the $\Z$ orbifold of the FZZT brane with Dirichlet boundary
conditions $\Theta^\alpha =0$. The orbifold operation
yields two fractional branes with opposite coupling
to the twisted sector, which we identify with
the $| \eta = -1, \xi = \pm 1 \rangle_B$ branes.  
This clearly agrees with the discretized expression \refb{Wx},
which cuts holes in the worldsheet with Dirichlet
b.c. for the $\Theta^\alpha$, and on which twist lines  can end.

\subsection{The ${\bf \eta = +1}$ brane}
\label{eta+}
The brane with $\eta = +1$ has so far been more puzzling, since it
does not appear to correspond to standard macroscopic loop operators
in the one- or two-matrix models \cite{chat3, SS}. In our formalism
(focussing on the $(2, 4)$ case)  there is a natural guess for what this brane should
correspond to,  namely the {\it other} natural choice of boundary 
conditions in the $\Theta^\alpha$ system:  Neumann boundary conditions. 
The most compelling piece of evidence for this proposal comes
from consideration of the open string spectrum between
Neumann and Dirichlet branes for the $\Theta^\alpha$ system,
which lies in the twisted sector  (see \refb{openchannel}) --
and hence has ``half-integer'' Liouville dressing ({\it i.e. }  $n$ {\it even} 
in the notations of section 4.2). 
This agrees with the fact that the open strings between $\eta = -1$ and $\eta = +1$ branes
are in the Ramond sector.  Since the identification of the $\eta = -1$ brane
with Dirichlet b.c. for $\Theta^\alpha$ is beyond any doubt,  
we seem to be forced to  identify the $\eta = +1$ brane with Neumann b.c. for $\Theta^\alpha$
in order to reproduce the correct open string spectrum.

\smallskip

In the matrix model, we can find an expression for the Neumann
macroscopic loop operator by appealing to the string theory
intuition.  We regard the matrix model as the open string
field theory on  Neumann ZZ branes  as in \cite{MV, KMS}. To a Neumann FZZT brane we add
 the open strings stretched between 
the Neumann FZZT brane and the $N$ Neumann ZZ branes.
Such open strings are expected to be fermionic \cite{MMSS},
and since we have $\Theta^\alpha$ Neumann b.c. on both branes, 
they will depend on the  $\Theta^\alpha$ zero modes. 
Thus we need to introduce a Grassmann odd open string field $\pi ( \theta, \bar \theta)$
in the fundamental of $U(N)$. In the supersymmetric
formulation we think of it as a new superfield.
We are led to the matrix model
\be
\int d {\bf \Phi}\, d \pi \; e^{-(S_{\bf \Phi} + S_\pi)}
\ee
with
\be \label{Sphib}
S_{\bf \Phi} =  \int d \theta d \bar \theta \; {\rm Tr}
( \partial_\theta {\bf \Phi} \partial_{\bar \theta} {\bf \Phi} 
 - a {\bf \Phi}^2 + {\bf \Phi}^3  )  \, ,
 \quad
  S_\pi =  \int d \theta d \bar \theta \; (\partial_\theta {\bf \pi}^\dagger \partial_{\bar \theta} {\bf \pi} + w  {\bf \pi}^\dagger {\bf \pi}  +  {\bf \pi}^\dagger {\bf \Phi} {\bf  \pi})  \, , 
  \ee
where $w$ is a mass parameter. In the Feynman diagram expansion, the new field introduces 
boundaries which are free to fluctuate in the $\Theta^\alpha$ directions. In the loop
gas picture, now there {\it will} be ghost lines attaching to the boundary. The $\pi$ superfield
can be integrated out, 
\be \label{Ndet}
\int d{\bf \Phi}\cdot \det(\Phi+w)^{-2}\cdot
\det\left((\Phi+w)^2-\bar{C}\frac{1}{\Phi+w}C-F\right)
\cdot e^{-S_{\bf \Phi}}.
 \ee
One can imagine to further integrate out $F$, $C$ and $\bar C$. This
would lead to a complicated expression involving a sum of multitrace combinations
of the lowest-component matrix $\Phi$. We leave a detailed
study of this new resolvant and of its double-scaling limit for future work. 

\smallskip
It would be nice to understand
in the framework of the continuum bosonic formulation
why  Neumann brane for $\Theta^\alpha$ 
should be a source of RR charge in the 
0A model, but not in the 0B model (see (\ref{Bbranes}, \ref{Abranes})).
At present we have no crisp answer. RR charge is related to the very subtle RR ground state,
which both in the supersymmetric and 
in the bosonic formulations precisely saturates the Seiberg bound.
A logical possibility is that while minimal
superstrings are perturbatively equivalent to bosonic
strings in the closed string sector, this equivalence
breaks down for D-branes,  which  
are non-perturbative objects.  This is a valid objection for ZZ branes.\footnote{We thank
N. Seiberg for raising this point.}
In the matrix model ZZ branes are related
to stationary points of the effective eigenvalue
potential, and the non-perturbative restriction on the range of eigenvalue integration will affect their 
spectrum. However, we expect that FZZT branes should have
a good description in the  bosonic formulation. 
Intuitively this is because FZZT branes can be continuously connected to
the closed string vacuum by sending the boundary cosmological constant to infinity, and in
that limit they admit an expansion in closed string operators. We hope that 
a better understanding of the matrix model \refb{Sphib} and of
its continuum limit will clarify this issue.

\subsection{Comparison with previous work}

We would like to make contact with previous discussions of boundary conditions for loop gas models 
\cite{KazKostov, KostovBoundary, Knew}. 
Loop gas models can be formulated
as  ``height'' models (see \cite{Knew} for a recent discussion),  
where configurations are specified by assigning a height variable 
to each vertex of a triangulated random surface (the heights of two
neighboring vertices differing at most by one unit).
The self-avoiding loops are the domain walls separating regions of the surface
of different heights.  The height model of \cite{Knew}
is rich enough to describe the (dicretized version of)
non-minimal matter of central charge $c_M \leq 1$
plus Liouville CFT of central charge $c_L = 26-c_M$. 
 In the continuum limit, the height variable of \cite{Knew}
goes over an anomalous free boson $X$  which
provides a Coulomb gas representation  of the matter CFT.
In the height model, the natural boundary conditions are ``Dirichlet'' (fixed height at the boundary)
and ``Neumann'' (free sum over different heights at the boundary). ``Dirichlet''
boundary conditions are equivalent to {\it not} allowing
the self-avoiding loops to touch the boundary. In the continuum limit,
they correspond to  Dirichlet boundary conditions
for the matter field $X$, and FZZT boundary conditions
for the Liouville BCFT.  The ``Neumann'' boundary conditions 
of \cite{KazKostov, KostovBoundary, Knew} correspond instead to 
summing over all possible configurations, with self-avoiding
loops allowed to end on the boundary.
 In the continuum limit, these boundary conditions go over Neumann boundary conditions
for the Coulomb gas field $X$, and {\it dual} FZZT boundary conditions for the Liouville \cite{Knew}. Dual FZZT boundary conditions mean
that the {\it dual} deformation
 $ e^{\phi/b}$ is turned on the worldsheet boundary as opposed to $   e^{b \phi}$.\footnote{
 Recall that our conventions are such that $b^2 = p/q$ (so $b$ can be either
 smaller or bigger than one according to the model), the boundary
 cosmological constant operator is always $e^{b \phi}$ and
 the dual boundary cosmological constant operator $e^{\phi/b}$.
 In \cite{Knew} different conventions are used, with $b <1$ by definition.}
In fact, we know from the  FZZT  boundary CFT \cite{FZZT} that
both  the $e^{b \phi}$ and $e^{\phi/b}$ deformations are
actually turned on in the exact solution, so in the Liouville sector the distinction
between ordinary and dual FZZT b.c. is immaterial (up to  relabeling  parameters);
however  in the matter sector,
the distinction between ``Neumann'' and ``Dirichlet'' b.c. for $X$ is significant.

\smallskip

The boundary conditions that we have identified
with the $\eta = -1$ FZZT brane, namely Dirichlet boundary
conditions for $\Theta^\alpha$, are clearly the same as the
``Dirichlet'' boundary conditions of \cite{Knew}.
However it is not at all obvious  that Neumann boundary conditions
for the $\Theta^\alpha$ system (and their discretized version \refb{Sphib}) are the same as
the ``Neumann'' boundary conditions of \cite{Knew}. 
The continuum limit of the height model
of \cite{Knew}  (specializing their formulas to $n  = -2$ in the dilute phase) 
gives a free anomalous boson $X$ with $c_M = -2$
coupled to Liouville. Clearly  ``Neumann'' boundary conditions  for 
$X$ have nothing to do with Neumann boundary conditions
for $\Theta^\alpha$. However this comparison with the formulas of
 \cite{Knew} is naive, because their model is different from ours in some crucial ways,
 even in the one-color case. The supersymmetric matrix model corresponds to a  $O(-2)$ loop gas
 model with {\it multicritical} quartic potential \cite{KS}; 
whereas the model of \cite{Knew} is based on a {\it triangulation} of the surface. Another
(probably related) point is that the $\Theta^\alpha$ CFT does not need screening charges, 
whereas the $X$ CFT does. It will be very interesting
to extend the formalism of \cite{Knew} to our case. It is
conceivable that a suitable modification of the loop equation approach of \cite{KazKostov, Knew} 
to ``Neumann'' boundary conditions may turn out to be equivalent to \refb{Sphib}, and provide a way to do 
concrete calculations.

\sectiono{Outlook}
\label{conclusions}

We have seen that the matrix models of minimal superstrings
have a very natural formulation as loop gas models on random
surfaces.  This language has many virtues.

\smallskip
First, it  clarifies the $\Z$ orbifold relation between the 0A and 0B matrix models, 
which amounts to going from an ordinary random surface
to a bicolored random surface.  Second, it gives
 in the continuum limit 
a formulation of minimal superstrings as bosonic string theories, to each order in the perturbative
expansion. The main news here are for the 0B cases, since the 0A cases
have no local
RR operators in the closed string spectrum and have
been known to be perturbatively equivalent to 
minimal bosonic theories \cite{chat3, cliffold, cliff1, cliff2}. In the 0B examples,
RR operators are in the twisted sector
of the $\Z$ orbifold and we have seen very explicitly how the loop gas model implements the sum 
over spin structures;
in the simplest case of $\hat c = 0$ this is the sum over spin
structures of the $\Theta^\alpha$ system.  One of the most
intriguing directions for future work is to see if
one can recover the loop gas model from a direct discretization
of the RNS worldsheet. The $\Theta^\alpha$ fields
seem vaguely reminiscent of gravitinos, in that RR operators
make them multi-valued  -- but they have, of course, the wrong spin.

\smallskip

Finally, and this may turn out to be the most practical aspect of our work,
loop gas models  appear to give a more transparent 
description of open string boundary conditions.
We find it very plausible that the $\eta = +1$ branes,
which are mysterious in other formulations,
correspond to Neumann boundary conditions for the $\Theta^\alpha$
system. Further work is needed to completely settle this issue.
A promising direction is a detailed study the new  macroscopic loop operator that we have 
proposed in section \ref{eta+}.  

\smallskip

Several extensions of the results of this paper can be contemplated. More
general $(p, q)$ minimal superstrings should be described
by two-matrix models, and their gauge-fixing
should lead to loop gas models. It would also be nice to see if a similar gauge-fixing 
procedure as described in this paper can give some insight into $\hat c = 1$ string theory.
Some progress in connecting  the $\hat c = 1$ Type 0A theory 
with the RNS formulation has been made in the interesting paper \cite{kapustin},
and it would be nice to understand the relation between their approach and
ours.

\smallskip

The  theme that underlies the whole subject is open/closed
duality. The doubled scaled matrix models dual to  non-critical
strings are interpreted as the effective open string field theories 
on a large number of decayed ZZ branes. In this paper we have seen a new sharp application of this
idea. The one- and two-color loop gas models are recognized
as the effective worldvolume theories on the appropriate
ZZ branes; it is satisfying to see that the
$\Z$ orbifold relating them is the usual  string theory \cite{DM}.
The {\it other} class of  Liouville branes, namely the FZZT branes,
offer an alternative route to open/closed duality \cite{GR}, which
we have not explored in this paper. The OSFT
on the FZZT branes of $(1, 2)$ minimal
bosonic string localizes to the  Kontsevich matrix model \cite{GR, MMSS}.
The speculation of \cite{GR} is that ``topological'' matrix models
{\it \` a la} Kontsevich are generically related to FZZT branes.   In the context
of minimal superstrings this raises many natural questions. We expect that the study of FZZT branes in
minimal superstrings will lead to interesting generalizations 
of topological matrix models. 

\smallskip

As usual with solvable models, the most important question
is which features of the exact solution generalize to more physical 
situations. Open/closed duality is 
certainly one such feature. Other lessons
specific to minimal {\it super}strings should be
sought in the study of exact RR backgrounds.
In the critical RNS string, there are two related difficulties with 
RR backgrounds:  they introduce cuts on the worldsheets and thus cannot be exponentiated 
in any obvious fashion; they break superconformal symmetry
and thus the very rules for string theory computations are unclear. 
In this paper we have seen that for minimal superstrings the second
difficulty can be circumvented by going to
an equivalent bosonic formulation. This
simplification may well be an artifact of the simplicity
of these models. However, the first problem
is still very much with us in the continuum bosonic formulation.
Remarkably, the matrix model manages to produce
an exact answer. A paradigmatic example
is the Ising model with external magnetic field.
Either as a continuum CFT, or as a spin
system on a regular lattice, this is a notoriously difficult
problem. The model becomes vastly
simpler once formulated on a random lattice \cite{ising};
it can be easily mapped  to a two-matrix model with asymmetric
potential, and solved exactly. Summing over triangulations before summing over the spin degrees 
of freedom  is the winning strategy. We can only
speculate about the implications for critical superstrings.
Perhaps the message here is that 
in trying to formulate RR backgrounds
on a fixed Riemann surface,  we are addressing 
a more difficult problem than we really need to solve.

\section*{Acknowledgments}

We are grateful to N. Seiberg for illuminating discussions and suggestions.
We are glad to acknowledge useful conversations with I. Klebanov, I. Kostov,  J. McGreevy,
D. Shih and H. Verlinde. DG and TT thank S. Minwalla very much for encouragement
to work on these topics and for collaboration at an early stage.

The work of DG and TT is supported in
part by DOE grant DE-FG02-91ER40654.
This material is partly based upon work (LR) 
supported by the National Science Foundation Grant No. PHY-0243680. Any opinions, findings, and
conclusions or recommendations expressed in this material are
those of the authors and do not necessarily reflect the views of
the National Science Foundation.

\appendix

\sectiono{Appendix: The ${\bf \Theta^\alpha}$ CFT}

In this appendix we review some basic facts 
about the ``sympletic fermions" CFT  \cite{Kausch1,  Kausch2,  Flohr, GK1, GK2, KoganWheater, 
KawaiWheater, BredthauerFlohr}
and elaborate on its connections with minimal superstrings. 

\subsection{Bulk theory}

$\Theta^+ (z, \bar z)$ and $\Theta^- (z, \bar z)$
are non-chiral, Grassmann odd fields of dimension zero,
governed by the free action
\be
\frac{1}{4\pi} \int d^2 z \, \epsilon_{\alpha \beta}\, \partial \Theta^\alpha 
\bar \partial \Theta^\beta \, .
\ee
Here $\alpha, \beta = \pm$,   $\epsilon^{\pm \mp} =\pm 1$,  $\epsilon_{\alpha \beta}
\epsilon^{\beta \gamma} = \delta_{\alpha}^{\,\gamma}$.  
Hermitian conjugation acts as $(\Theta^+)^\dagger = \Theta^-$.
This CFT differs from the more familiar $\eta \xi$ system
in the treatment of the zero modes. One can identify
$\xi(z) + \bar \xi(\bar z) = \Theta^+(z, \bar z)$,
$\eta(z) = \partial \Theta^- (z, \bar z)$ and $\bar \eta(\bar z) = \bar \partial \Theta^- (z, \bar z)$,
but there is a mismatch in the zero mode sector. While the $\eta \xi$ system has two 
{\it chiral} zero modes (one for the chiral field $\xi$ and one for the antichiral
field $\bar \xi$), the $\Theta^\alpha$ system
has two {\it  non-chiral zero} modes,  one for $\Theta^+$ and one for $\Theta^-$.

\smallskip
Let us first consider the ``untwisted'' theory, with periodic fields
$\Theta^\alpha (z, \bar z) = \Theta^\alpha ( e^{2 \pi i} z, e^{-2 \pi i} \bar z)$. The  
mode expansion  reads
\be
\Theta^\alpha ( z, \bar z) = \theta_0^\alpha +\chi_0^\alpha \ln |z|^2
+ i \sum_{n = -\infty\, n \neq 0 }^\infty 
\left( 
\frac{\chi_n^\alpha }{n  z^n}+ \frac{\bar \chi_n^\alpha }{n \bar z^n}   
\right) \, .
\ee 
The modes obey the obvious canonical anticommutation relations
\be
\{     \chi_0^\alpha ,  \theta_0^\beta \} =   \epsilon^{\alpha \beta}  \, ,
\quad  \{    \chi_m^\alpha,   \chi_n^\beta  \} = m \epsilon^{\alpha \beta}\delta_{ m+n, 0}
\, , \quad   \{    \bar \chi_m^\alpha,  \bar \chi_n^\beta  \} =m \epsilon^{\alpha \beta}\delta_{n + m, 0}\,.
\ee
Notice that in correspondence with the two non-chiral zero modes $\theta_0^\alpha$, there are 
 two canonically conjugate non-chiral momenta $\chi_0^\alpha $,
{\it i.e.} one has the identification $\chi_0^\alpha \equiv \bar \chi_0^\alpha$,
which is required by locality.  The Fock space is obtained
by acting with the creation operators $\chi^\alpha_{-n}$ and $\bar \chi^\alpha_{-n}$
$ (n > 0)$, on the space of ground states, which is spanned by
\be \label{vacua}
| \Omega \rangle  \,  ,\qquad  | \theta^\alpha \rangle \equiv  \theta_0^\alpha | \Omega\rangle
\, , \qquad |\omega \rangle \equiv  \theta_0^+ \theta_0^- | \Omega \rangle \,.
\ee
Here $| \Omega \rangle$ denotes the $SL(2, {\bf C})$ vacuum. This system
provides  the simplest realization of a logarithmic CFT. The states
$|\Omega \rangle$ and $| \omega \rangle$ form a logaritmic pair, spanning a two dimensional
Jordan block for $L_0$,
\be  
L_0 | \omega \rangle = \bar L_0 | \omega \rangle = | \Omega \rangle \,,
\quad L_0 | \Omega \rangle = \bar L_0 |\Omega \rangle = 0 \,.
\ee
Let us also record the inner products 
\be
\langle \Omega | \Omega \rangle = \langle \theta^\alpha | \Omega \rangle =
\langle \theta^\alpha  | \omega \rangle = 0 \,,  \quad \langle \Omega | \omega \rangle =
 1 \, ,  \quad
\langle \theta^\alpha | \theta^\beta \rangle = \epsilon^{\alpha \beta} \,,\quad
\langle \omega | \omega \rangle = \kappa \, .
\ee
Here $\kappa$ is an arbitrary normalization constant, corresponding to the freedom
to redefine $| \omega \rangle \to | \omega \rangle + \alpha | \Omega \rangle$.
If we stick to the definition in \refb{vacua}, then $\kappa = 0$.

\smallskip

The path-integral  is zero unless one saturates the 
fermionic  zero modes. The simplest non-zero correlators
involve a single insertion of the logarithmic identity $\omega$,
\be \label{omegacorr}
\langle   \omega (z, \bar z)  \,{\cal F}_1   (w_1, \bar w_1)  \cdots 
  {\cal F}_n   (w_n, \bar w_n)\rangle \, ,
\ee
where $ {\cal F}_k   (w_k, \bar w_k)$ is an  arbitrary local operator
containing no zero modes. Such a correlator does not depend on the position
of the  $\omega$ insertion,  since taking a derivative with respect to $z$ gives a correlator that has
unbalanced zero modes and is thus zero. On the sphere,
a correlator with a single $\omega$ insertion factors
into a  holomorphic rational function of the coordinates
$w_i$ times a antiholomorphic rational
function of $\bar w_i$. Correlators involving
more than one $\omega$ insertion contain logarithmic
terms  $\log |w_i - w_j|$, but they will not be important for us.
 Notice that even on higher genus Riemann surfaces one insertion
 of $\omega$ is sufficient to saturate the zero modes. Moreover, on a surface of arbitrary
 genus,  the path  integral in the non-zero mode sector will vanish unless   the number of $\alpha = +$ insertions equals  the number
 of $\alpha = -$ insertions.
 This is in contrast with the $\eta \xi$ system, where on a surface
 on genus $g$ one needs to saturate  one chiral $\xi$ zero mode
 (the constant mode) and  $g$ chiral $\eta$ zero modes (corresponding to the $g$ holomorphic one-forms);
and similarly one $\bar \xi$ and $g$  $\bar \eta$ antichiral zero modes.
 The torus partition function with periodic boundary
 conditions and one insertion of the logarithmic identity reads
 \be
 {\rm Tr} \left[(-1)^F \, \omega \, q^{L_0 + 1/12} \bar q^{\bar L_0+ 1/12} \right]=
 2 \pi \tau_2 | \eta(\tau)|^4 \, ,
 \ee
which is just the inverse of the partition function for two free bosons, and is  clearly modular
invariant.

 \smallskip
 
 For our purposes, the $\Theta^\alpha$ theory with periodic fields
 is the matter CFT with $c_M = -2$ that  enters in the construction
 of the $(1, 2)$ bosonic string. As usual, the string theory is defined
 by coupling the matter CFT with Liouville CFT of $c_L =
 26 - c_M = 28$ and the $bc$ ghosts of $c_G = -26$. As observed long ago
 in \cite{distler}, the connection with topological gravity is provided by
 taking the Liouville CFT to have  $\mu = 0$ and performing
  the bosonization\footnote{ The arguments of \cite{distler} are actually phrased
  in terms of $\eta \xi$ system.}
  \beq 
  \label{boson}
 \beta = \partial \Theta^+ e^{\frac{\phi}{\sqrt {2}} } \, ,\quad
\gamma = \partial \Theta^- e^{-\frac{\phi}{\sqrt {2}} } \, ,
 \eeq
 where $\phi$ is the Liouville field. The bosonic ghosts $\beta \gamma$ have central charge 26 and
 are the superpartners  of the fermionic ghosts $b c$ in the
 topological gravity multiplet of \cite{LPW}.  The bosonization
 \refb{boson} does not involve the $\Theta^\alpha$ zero modes.
 If we want the path integral over $\phi$ and $\Theta^\alpha$ to reproduce the $\beta \gamma$
path integral, we should simply insert a single logarithmic identity $\omega$ 
in every correlator, as in \refb{omegacorr}.  Similarly, 
since the Liouville path integral for $\mu = 0$  is proportional to the (infinite)
volume ${\rm Vol}(\phi_0)$ of the Liouville zero mode, 
in making contact with the $\beta \gamma$ system we need to  
divide out by  ${\rm Vol}(\phi_0)$.\footnote{The discrete analog
of dividing out by  ${\rm Vol}(\phi_0)$  is to divide out by
the logarithmic divergence $|\log {\Delta}|$ found in the matrix model
(where $\Delta$ is the bare cosmological constant),  as is done in  \cite{KW2}.}
 With these rules,
the simplest non-vanishing correlator is the sphere
amplitude of three cosmological constant operators
${\cal O}_1 = c \bar c e^{\sqrt{2}\phi}$ (with the implicit
extra insertion of $\omega$ at some arbitrary point on the surface). 
This correlator is of course just a constant, in accord with the 
fact that the topological gravity \cite{Wit} 
partition function at genus zero is proportional $\mu^3$.

 \smallskip
 
 The $\Z$ orbifold $\Theta^\alpha \to - \Theta^\alpha$
 projects out all odd states and introduces the twisted
 sector with antiperiodic boundary conditions
  $\Theta^\alpha (z, \bar z) = -\Theta^\alpha ( e^{2 \pi i} z, e^{-2 \pi i} \bar z)$.
 There is a unique twisted sector ground state $| RR \rangle $, with dimension  $(-1/8, -1/8)$.
The full torus partition function
 of the $\Z$ orbifold,  obtained by summing over spin structures, can be written as \cite{Kausch1}
 \be
 \frac{1}{|  \eta(\tau)|^2} \sum_{m,n = -\infty}^\infty q^{\frac{1}{4} (\sqrt{2} m + n/\sqrt{2})^2}
 \bar q^{\frac{1}{4} (\sqrt{2} m - n/\sqrt{2})^2} \, ,
 \ee
 which one recognizes as  identical to the 
 partition function of a compact boson at radius $R = \sqrt{2 \alpha'}$. The
 $\Z$ orbifold theory is identical to the ``triplet model'' of \cite{GK1, GK2}, which
 was originally introduced  by an algebraic construction involving
 a triplet of  $W_3$ generators. In the explicit free field representation provided
 by the symplectic fermions, the $W_3$ generators are the spin 3 operators 
 $\partial^2 \Theta^+ \partial \Theta^+$,
 $( \partial^2 \Theta^+ \partial \Theta^- + \partial^2 \Theta^- \partial \Theta^+)$ and
  $\partial^2 \Theta^- \partial \Theta^-$. The OPE of these 
operators close with the stress tensor to form an enlarged chiral algebra.  The theory is rational 
with respect to this extended algebra.

 \subsection{Boundary states}
 
 Open string boundary conditions for symplectic fermions have been studied
 in \cite{KoganWheater, KawaiWheater, BredthauerFlohr}.  Boundary states preserving the $W$-algebra
 are either Neumann or Dirichlet,
 \be \label{cond}
 (\chi_m^\alpha - \bar \chi_{-m}^\alpha) | {\rm Dir} \rangle = 0\, ,
 \quad  (\chi_m^\alpha + \bar \chi_{-m}^\alpha) | {\rm Neu} \rangle= 0\,  , \quad \forall m \geq 0 \, ,
 \ee
 with the understanding that $\chi_0^\alpha \equiv \bar \chi_0^\alpha$.
 In the untwisted theory, these conditions are solved by 
 \ben \label{states}
 | {\rm Dir} \rangle & =& {\cal N}_D
 \exp \left(  {\sum_{k > 0 } \frac{1}{k} \chi_{-k}^- \bar \chi_{-k}^+  +  \frac{1}{k} \bar \chi_{-k}^-  
 \chi_{-k}^+     }\right)  | 0 \rangle_D\, ,  \\
  | {\rm Neu} \rangle   & = & {\cal N}_N
 \exp \left(  {\sum_{k > 0 } -\frac{1}{k} \chi_{-k}^- \bar \chi_{-k}^+    -\frac{1}{k} \bar \chi_{-k}^-  
 \chi_{-k}^+     }\right)
  | \Omega \rangle \, . \nonumber
 \een
 Here $| 0 \rangle_D$ indicates any  of the vacua
 in \refb{vacua}, since for the Dirichlet case the $m = 0$ condition in \refb{cond}
is trivially implied by locality ($\chi_0^\alpha \equiv \bar \chi_0^\alpha$). In fact
it is natural to assemble the four possible Dirichlet boundary states into
the linear combination
 \be \label{Dgen}
|  {\rm Dir}  (\rho^\alpha ) \rangle  = {\cal N}_D \, \exp({ \epsilon_{\alpha \beta}\rho^\alpha \chi_0^\beta}  )
\; \exp \left(  {\sum_{k > 0 } \frac{1}{k} \chi_{-k}^- \bar \chi_{-k}^+  +  \frac{1}{k} \bar \chi_{-k}^-  
 \chi_{-k}^+     }\right)  
   | \omega \rangle\, ,
\ee
where $\rho^\alpha$ are Grassmann numbers. Since
 $\theta_0^\alpha |   {\rm Dir}  (\rho^\alpha ) \rangle = \rho^\alpha  |   {\rm Dir}  
(\rho^\alpha ) \rangle$, this  is interpreted as the Dirichlet brane at ``position'' $\Theta^\alpha = 
\rho^\alpha$.  In the $\Z$ orbifold theory, the Dirichlet branes
built on $| \theta^\alpha \rangle$ are projected out. Two new states
satisfying \refb{cond} are built on the twisted sector ground state,
\be \label{mustates}
\exp \left(  {\sum_{k > 0 } \pm \frac{1}{k} \chi_{-k}^- \bar \chi_{-k}^+  \pm  \frac{1}{k} \bar \chi_{-k}^-  
 \chi_{-k}^+     }\right)  | RR \rangle \, ,
\ee
where the plus (minus) signs correspond to Dirichlet (Neumann) boundary conditions.
(Here   $\chi_k^\alpha$, $\bar \chi_k^\alpha$ are half-integer moded.)
The consistent boundary states obeying Cardy conditions are specific linear
combinations of the states in \refb{states} and in \refb{mustates}, we refer
to \cite{KawaiWheater, BredthauerFlohr} for details.  (The construction of the boundary states
is somewhat more subtle than in the paradigmatic minimal model example,
because of the non-trivial Jordan block structure of some of the
representations, but can nevertheless be carried out). 

\smallskip

The essential point
to emphasize here is that the open string spectrum
between a D-brane with Dirichlet boundary conditions and a D-brane
with Neumann boundary conditions lies in the twisted sector. Consider
the cylinder amplitude 
\be
\langle  {\rm Neu} | (\tilde q^{1/2})^{L_0 + \bar L_0 + 1/6}  |{\rm Dir }(\rho^\alpha ) \rangle =  
 \frac{{\cal N}_D {\cal N}_N}{4}\,  \tilde q^{1/12}
\prod_{n = 0}^\infty ( 1 + \tilde q^n)^2= 
 \frac{{\cal N}_D {\cal N}_N}{2}\,
\frac{\theta_2 (\tilde \tau)}{ \eta(\tilde \tau)} \,. 
\ee
Modular transformation to the open channel gives (with ${\cal N}_D {\cal N}_N = 2$) 
\be \label{openchannel}
\frac{\theta_4 (\tau)}{ \eta(\tau)} =  q^{c/24}   q^{-1/8} \prod_{n=0}^\infty  (1 - q^{n + 1/2})^2 \,.  
\ee
Here $c \equiv -2$.  We recognize the spectrum of
open strings obtained acting with half-moded oscillators 
$\{ \chi^\alpha_{-k} \;| \;  \alpha = \pm , \,  k = 1/2, 3/2, \dots \}$
on the twisted vacuum.

\begingroup\raggedright

\endgroup

\end{document}